# Path toward manufacturable superconducting qubits with relaxation times exceeding 0.1 ms

J. Verjauw[1,2], R. Acharya[1,3], J. Van Damme[1,3], Ts. Ivanov[1], D. Perez Lozano[1,2], F. A. Mohiyaddin[1], D. Wan[1], J. Jussot[1], A. M. Vadiraj[1], M. Mongillo[1], M. Heyns[1,2], I. Radu[1], B. Govoreanu[1], A. Potočnik*[1]

[1]**Imec, Kapeldreef 75, Leuven, B-3001, Belgium**
[2]**Department of Materials Engineering (MTM), KU Leuven, Leuven, B-3000, Belgium**
[3]**Department of Electrical Engineering (ESAT), KU Leuven, Leuven, B-3000, Belgium**

As the superconducting qubit platform matures towards ever-larger scales in the race towards a practical quantum computer, limitations due to qubit inhomogeneity through lack of process control become apparent. To benefit from the advanced process control in industry-scale CMOS fabrication facilities, different processing methods will be required. In particular, the double-angle evaporation and lift-off techniques used for current, state-of-the art superconducting qubits are generally incompatible with modern day manufacturable processes. Here, we demonstrate a fully CMOS compatible qubit fabrication method, and show results from overlap Josephson junction devices with long coherence and relaxation times, on par with the state-of-the-art. We experimentally verify that Argon milling - the critical step during junction fabrication - and a subtractive etch process nevertheless result in qubits with average qubit energy relaxation times $T_1$ reaching 70 μs, with maximum values exceeding 100 μs. Furthermore, we show that our results are still limited by surface losses and not, crucially, by junction losses. The presented fabrication process therefore heralds an important milestone towards a manufacturable 300 mm CMOS process for high-coherence superconducting qubits and has the potential to advance the scaling of superconducting device architectures.

# INTRODUCTION

Superconducting circuits have emerged as a leading candidate for realizing a scalable quantum computing platform. The improvements in qubit coherence times [1] and gate fidelities [2–5] enabled the demonstration of quantum simulators [6,7], small-scale quantum algorithms [8,9] and even the elusive demonstration of quantum supremacy [10,11]. State-of-the-art coherence times routinely reach 50 µs, with specific cases exceeding 100 µs [12–15]. These have been exclusively fabricated using aluminium (Al) double-angle evaporation and lift-off techniques on sapphire or high resistivity silicon (Si) substrates. Double-angle evaporated junctions enabled the fabrication of Noisy Intermediate Scale Quantum (NISQ) processors with an intermediate scale number of qubits [10,11], only recently exceeding 100 [16]. To further advance the technological state-of-the-art, the design and device fabrication inevitably become increasingly complex [17,18]. This imposes tighter constraints on the design parameters of qubits, readout resonators, on-chip filters [19,20], and tunable couplers [21,22]. To facilitate this upscaling, versatile, reliable, and reproducible fabrication processes are needed.

The large-scale implementation of superconducting qubits is inherently hindered by the variability in Josephson energy of the double-angle evaporated junctions [23,24]. This is primarily a consequence of the variability in fabricated junction area, which results from the angle dependence across the wafer during metal deposition [25–27]. Another limitation is that polymer masks that are typically used during processing restrict the thermal budget. This further limits the choice of superconductor, limiting the potential optimization space for qubit improvement [28,29]. Furthermore, the fabrication itself requires dedicated evaporation tools with tilt capability. Double-angle evaporation with a required lift-off step introduces resist contaminations [30] and reduces reproducibility in larger diameter wafers [25]. This fabrication technique is therefore considered incompatible with advanced complementary metal-oxide semiconductor (CMOS) manufacturing, where large-scale integration of devices is instead generally based on subtractive etch, sputtering deposition and advanced optical lithography [31]. To overcome these limitations, alternative junction fabrication techniques are being actively investigated.

Several alternatives to double-angle evaporated junctions exist, including finMET [32], trilayer [31,33] and overlap [11,34–38] Josephson junctions. These are compatible with manufacturable CMOS processes. The advantages of overlap junctions compared to other alternatives are the reduced structural complexity and therefore a smaller number of fabrication steps. This reduces the processing-induced sources of loss and parameter variability. Recent work has shown promising overlap junction qubit performance, with energy relaxation $T_1$ times up to 80 µs [34,38,39]. However, these qubits were still created using a CMOS-incompatible lift-off process. Qubits with state-of-the-art coherence times, fabricated with a fully manufacturable CMOS process, *i.e.*, without the use of lift-off techniques, sapphire or high resistivity (above ~10 kΩ·cm) Si substrates, are yet to be demonstrated.

Overlap junctions have two electrodes that are defined in two patterning cycles, with a vacuum break in between that results in the uncontrolled growth of native metal oxide. After the first cycle, the native oxide needs to be removed by *in-situ* argon (Ar) milling to enable controlled subsequent junction oxidation. This step is critical, since Ar milling has been linked to superconducting device performance degradation [15,40,41]. Strong Ar milling can lead to amorphized Si [42] and sapphire [30] substrate layers which were found to limit resonator quality factors and qubit coherence times. The milling has also been reported to introduce additional loss at metal-metal interfaces [15]. This indicates that the Ar milling could compromise the junction's integrity and composition. However, the impact of Ar milling on the junction performance is still unexplored, and at present it is unknown if it inherently limits the overlap junction qubit lifetimes.

In our work, we perform an in-depth characterization of Al overlap junctions, fabricated with wet and dry subtractive etch processes. Intrinsic Si substrate coupons with 3 kΩ·cm (3k) and 20 kΩ·cm (20k) specific resistivity are used from 300 mm and 100 mm wafers, respectively. Results for fixed frequency transmon qubits with overlap junctions show time-averaged energy relaxation $T_1$ times of 50-70 µs, with individual measurements surpassing 100 µs, and average Pauli gate fidelity of 99.94%. By comparing quality factors ($Q = 2\pi T_1 f$) of qubits with different capacitor geometries, we find that the total loss ($1/Q$) scales with the metal-air surface oxide participation ratio, suggesting that the qubits are limited by the surface losses residing at the capacitor pads instead of the junction. The dominant loss likely originates from the subtractive-etch-specific sidewall residues and native surface oxides. Our study reveals that the overlap junctions do not limit the $T_1$ coherence time up to at least 103 µs for qubits at 3 GHz.

# Results

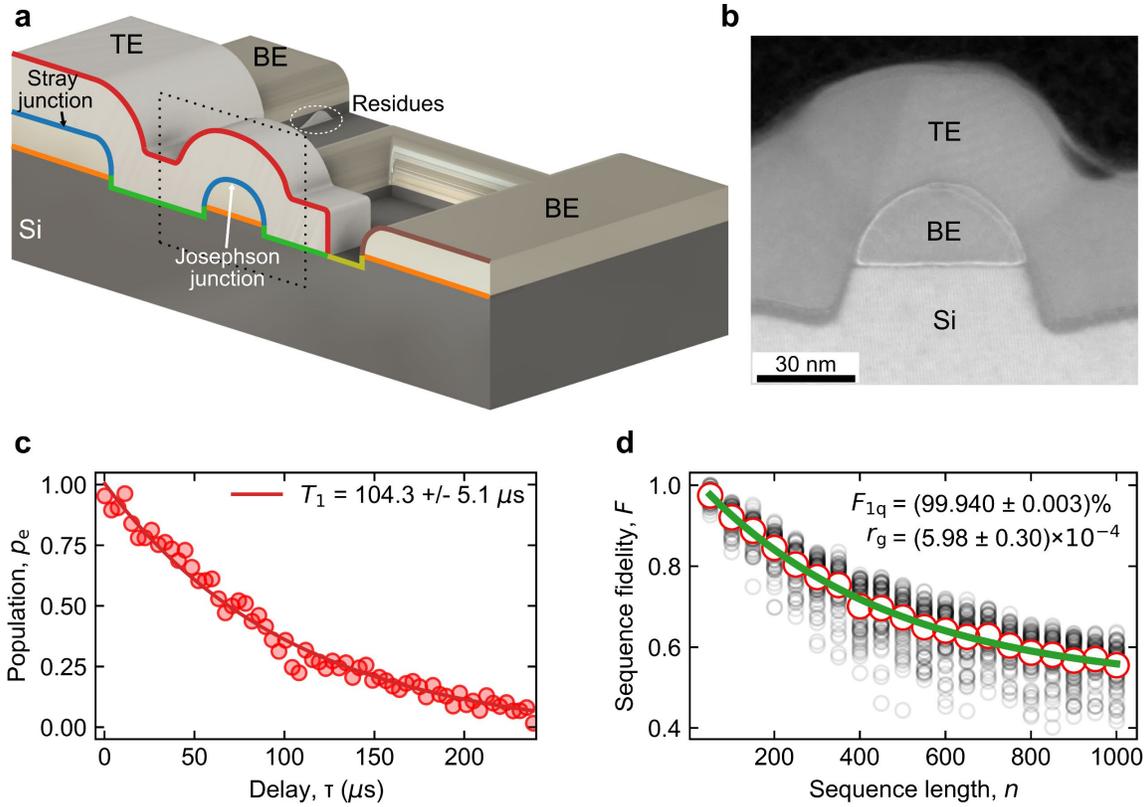

**Fig. 1**: Overlap junction overview and transmon qubit performance. **a** Cross-sectional illustration. The overlap between the bottom electrode (BE) and top electrode (TE) defines the Josephson junction and the (parasitic) stray junction. Sidewall residues can be present due to subtractive etching steps, which is discussed later. Si-BE, Si-TE and junction interfaces are indicated in orange, green and blue, respectively. The BE-air, TE-air and Si-air top-surface interfaces are depicted in brown, red and yellow, respectively. **b** High-angle annular dark-field-scanning tunnelling electron microscope (HAADF-STEM) image of a wet-etched BE junction cross-section indicated by the dotted line in **a**. Ar-milling induced amorphous Si layer is visible below the TE layer (green interface in **a**). **c** Qubit energy relaxation measurement with $T_1 = 104.3 \pm 5.1$ μs. **d** Average gate fidelity $(99.940 \pm 0.003)\%$ and average error per gate $(5.98 \pm 0.30) \times 10^{-4}$ measured with randomized benchmarking.

## Overlap junction fabrication and room temperature characterization

The process used to fabricate transmon qubits with overlap Josephson junctions is based on previously published work [34,39] and is described in detail in the Methods section and Supplementary Fig. 1. A 70 nm Al film is sputtered on a hydrofluoric (HF) acid-cleaned, high-resistivity Si substrate. Qubit capacitor pads, readout resonators, ground plane and overlap junction bottom electrode (BE) are defined simultaneously using either a dry or wet subtractive etch. Fabricating these structures in a single layer minimizes the number of interfaces. Next, native oxides from the entire surface including BE are removed using an optimized Ar milling step inside the deposition tool. This step impacts the BE-air, Si-air and junction interfaces. Following the native oxide removal, a controlled dynamic oxidation is performed to create the junction barrier. Next, a 50 nm thick top Al layer is deposited. In the final step, the top electrode (TE) is patterned with a subtractive dry etch. The targeted overlap junction areas are in the range of 0.03 - 0.07 μm$^2$. In addition to the overlap junction, the TE forms a large stray junction (~10 μm × 20 μm) used as galvanic contact to the qubit's capacitor pad (Fig. 1a: blue interface). A scanning tunnelling electron microscope (STEM) image of a fabricated overlap junction is shown in Fig. 1b. Devices fabricated with the described process exhibit record-high qubit energy relaxation times for overlap Josephson junction qubits, exceeding 100 μs (Fig. 1c). Furthermore, randomized benchmarking [43] yields average single qubit gate fidelities of $F_{1q} = (99.940 \pm 0.003)\%$ which are very close to the coherence limit of $F_{1q,inc} = 99.96\%$ (see Methods section and Fig. 1d). The uncertainty on the values represents the standard deviation (SD) on the fit.

We perform both wet and dry subtractive etching to define the BE. Wet etch processes are selective to Si and leave the substrate's top surface intact. It can be argued that they yield better qubit coherence times [14] and resonator quality factors [44] compared to dry etched devices, as the latter can be affected by an increase in Si substrate roughness and sidewall residues [14]. Nevertheless, recent demonstration of record-high relaxation times of dry etched qubits [12] indicates that these problems can be mitigated. Dry etch processes also provide better junction dimension control and reproducibility, making it the preferred technique in state-of-the-art CMOS processing [31].

To characterize fabrication performance, resistances of test Josephson junction arrays with different junction areas are measured at room-temperature. Arrays consist of either 30 or 60 individual test junctions and junction areas range between 0.03 and 0.125 $\mu m^2$, as determined from SEM images. The optimized fabrication process results in 99.8% junction yield for 3k dry etch samples. The resistance's relative standard deviations (RSD) range between 2-5% (Supplementary Fig. 2), which is comparable to the state-of-the-art double-angle evaporated Josephson junctions [23–25,45]. The reported RSD is notably lower than RSD estimated for previously reported overlap junctions [34]. Based on RSD analysis [24], we conclude that the junction parameter variation is dominated by the variation in junction area (Supplementary Fig. 2).

**Overlap Josephson junction qubit performance**

High coherence overlap junction qubits are characterized at 10 mK in a conventional dilution-refrigerator setup (see Methods Section and Supplementary Fig. 3). Double-pad transmon qubits (TM) show typical $T_1$ = 59 μs, Ramsey decoherence time of $T_2^*$ = 63 μs and spin-echo decoherence time of $T_{2e}$ = 59 μs (Fig. 2a). We extract a mean of $T_1$ = 58 μs, with observed values reaching up to 100 μs, from repeated measurements over a span of 11 hours for a dry etched TM qubit fabricated on a 3k Si substrate (Figs. 1c, 2b). The notable $T_1$ variation has been previously attributed to the presence of fluctuating TLS defects [46,47] and quasiparticle tunnelling through the Josephson junction [48]. We note that our reported record values surpass previously reported relaxation times for overlap junction qubits [34,38,39].

The impact of etching processes and substrate resistances on the qubit lifetimes is studied next. We collect $T_1$ statistics from devices fabricated with different process combinations: wet etch on 20k wafers (wet-20k), dry etch on 20k wafers (dry-20k) and dry etch on 3k wafers (dry-3k). The transmon qubits consistently achieve average $T_1$ lifetimes between 50-70 μs. The violin plot (Fig. 2c) represents kernel distribution estimates from measured $T_1$ times for each device. We observe no significant difference between the etch processes and wafer resistivities. This agrees with SEM images, where no visual difference is observed between junctions fabricated by the wet and dry etch processes (Supplementary Fig. 4). The $T_1$ independence on the substrate resistivity also agrees with our estimation of negligible microwave losses from similar 3k substrates ($Q_i$ ~ 8 M) of previous resonator measurements [49], setting an upper limit of the qubit $T_1$ > 400 μs at 3 GHz. Within the measurement uncertainty, the qubit lifetimes therefore appear not to be compromised by the dry etch manufacturable process on 3k wafer substrates.

To further investigate the sources of loss in overlap junction qubits, we design and fabricate two additional Xmon type qubits [50] (XM1, XM2) with reduced gaps between the centre electrode pad and the ground plane (Supplementary Fig. 5). By engineering the geometry and gap size, we control the participation ratios (fraction of the device's total electric field energy stored in the volume of lossy dielectric materials) of surface interfaces such as the substrate-air (SA) and metal-air (MA) interface [50–52]. For each design, we estimate participation ratios with electrostatic simulations (see Methods section and Supplementary Fig. 5). The Josephson junction dimensions (Supplementary Table 1), and its associated interface participation ratios are similar and uncorrelated with the qubit design. This allows us to distinguish between losses originating from the junction and its vicinity (junction losses) and from the qubit capacitor pads (capacitor losses).

For a proper comparison of qubit lifetimes at different transition frequencies, Q-factors, calculated from mean qubit relaxation times and their transition frequency ($Q = 2\pi T_1 f$) [47,53,54], are compared to the MA participation ratio of different qubit designs. The MA participation ratio was chosen since its interface was found to dominate microwave loss [52] and generally scales with other interface participation ratios (Supplementary Table 2). We find that the reciprocal of the qubit's Q-factors (the qubit's total loss) scales with the qubit MA participation ratio, as indicated by the linear fit of the qubit data in Fig. 2d. The linear fit agrees with the on-chip coplanar waveguide resonator (RES) datapoints (open symbols in Fig. 2d). This confirms that qubits and resonators share a common dominant loss source, associated with surface interfaces. For qubits, the dominant loss therefore originates from the interfaces at capacitor pads.

From the $y$-axis intercept of the linear fit, we can extract a hypothetical residual loss of $1/Q = 0.56 \times 10^{-6}$, corresponding to a postulated, junction-limited $T_1 \approx 103 \pm 40$ μs for a qubit transition frequency of 3 GHz. The uncertainty is estimated from the 95% confidence bound. The residual loss is associated with the junction and constitutes a combination of losses coming from the Josephson junction and interfaces at its vicinity [42,55], the stray junction [56], or quasiparticle tunnelling [48].

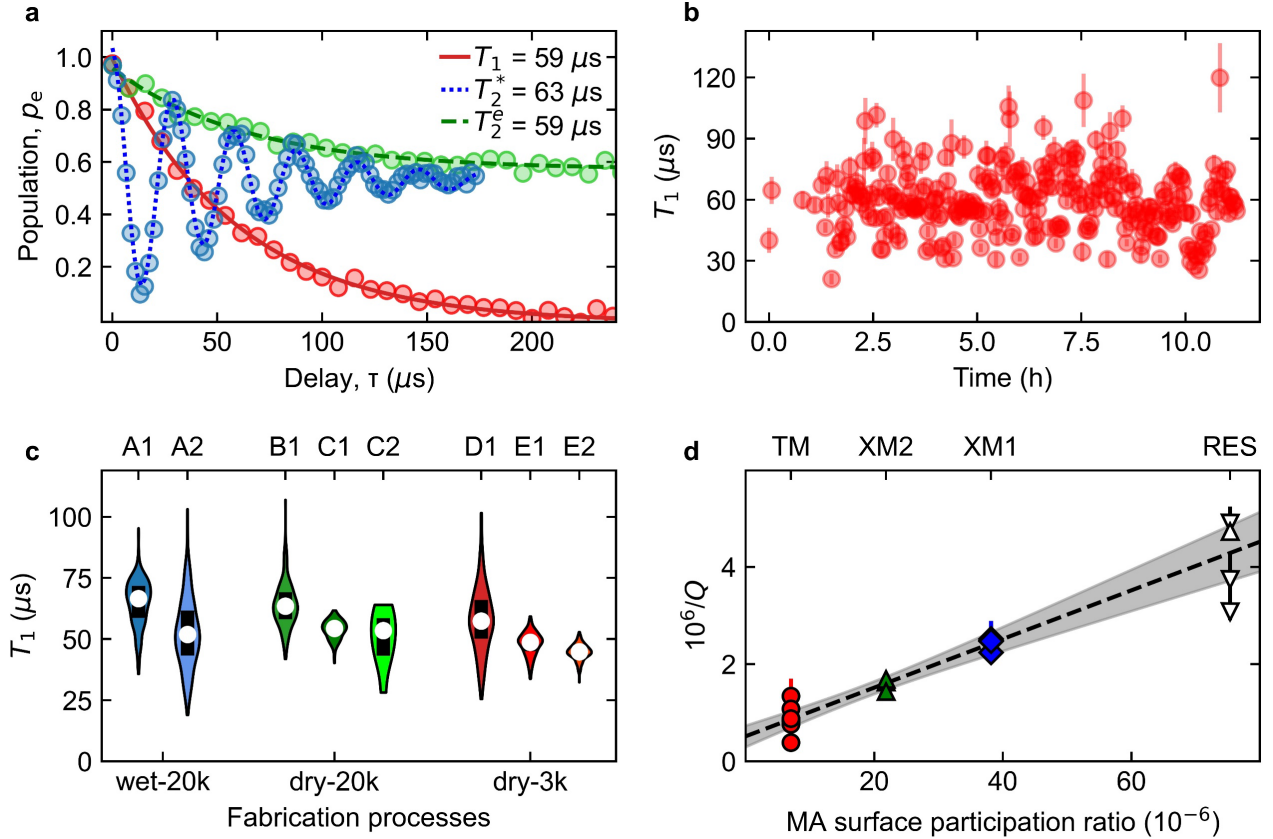

**Fig. 2**: Qubit coherence time characterization. **a** Qubit population (device C1) during typical decay time measurements. Line fits are shown for the qubit's lifetime $T_1 = 59.6 \pm 1.4$ μs, and coherence times $T_2^* = 63.8 \pm 3.9$ μs (Ramsey) and $T_{2e} = 59.3 \pm 2.0$ μs (spin-echo). **b** Fluctuation of the qubit's energy relaxation time for a period of 11h (device D1). Error bars represent the $T_1$ SD from the line fit. **c** Qubit $T_1$-statistics for several transmon devices, grouped for different process variations. Device names are shown on top. A violin plot is used to visualize the distributions and white dots indicate the mean value. Black bars mark the first and third quartile. Within the measurement uncertainty, no conclusive difference between etch processes and wafers resistivities is observed. **d** Inverse device quality factors as a function of MA surface participation ratio. Device types are shown on top. The grey area represents the 95% confidence bounds of the fit to the qubit data. Markers denote the mean $Q$-factor for each device. Error bars indicate the population's SD (or the fit's SD in case of single measurements). Upward and downward empty triangles depict wet and dry etched RES devices, respectively. RES data reaffirms that both etch processes result in comparable device performance. Device details are presented in Supplementary Table 1.

### Interface characterization

We attempt to correlate the origin of the qubit losses and the possible effect of Ar milling by examining interface morphology and atomic compositions with high-resolution cross-sectional Transmission Electron Microscopy (TEM) and Energy-Dispersive X-ray Spectroscopy (EDS), respectively. The characterization results depicted in figure 3 are of a dry-etched qubit device (see Methods section for details).

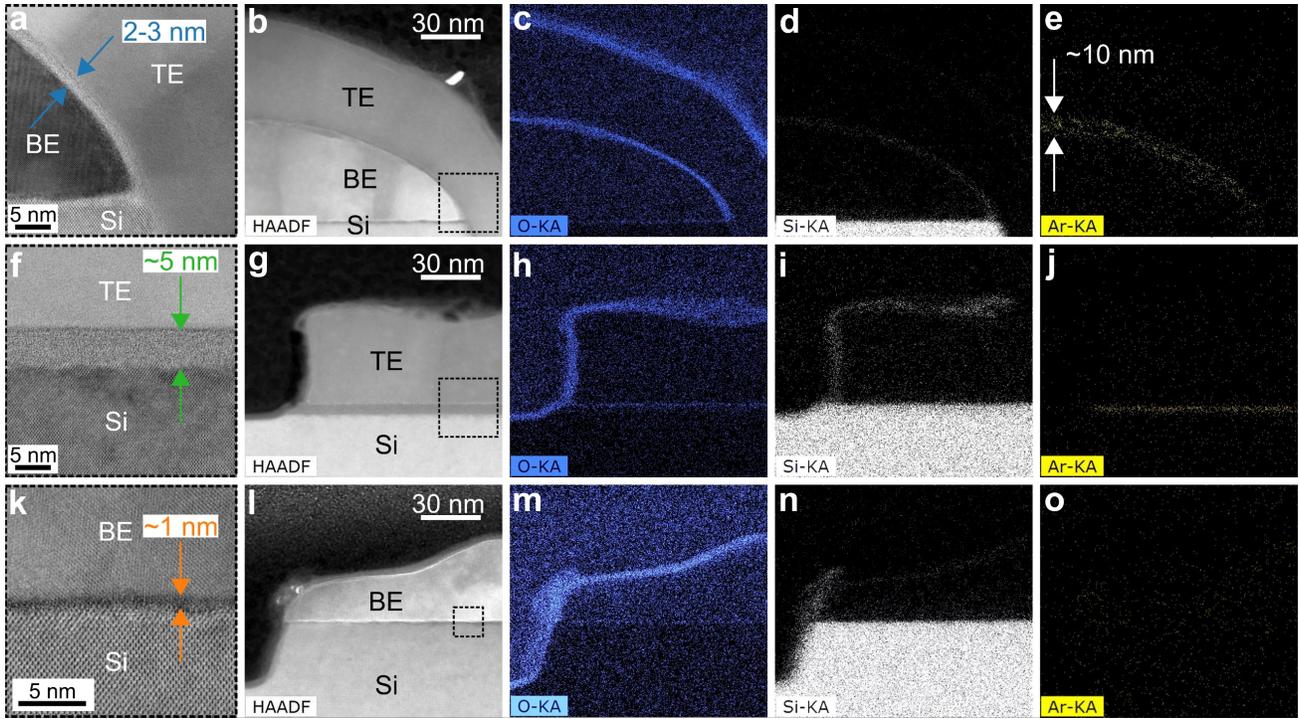

**Fig. 3**: EDS results of the O/Si/Ar content for dry etched sample interfaces. For each row, the position of the high-resolution annular bright-field-scanning transmission electron microscope (ABF-STEM) image in the first column is indicated by the dashed square in the HAADF-STEM image in the second column. The four rightmost columns have the same scale. **a-e**: Josephson junction profile. In addition to O, traces of Ar and Si atoms are found in the junction, likely due to Ar milling during native oxide removal. **f-j**: Si-TE, TE-air and Si-air interfaces. The amorphous Si layer (containing Ar and O similar to the junction) is encapsulated below the TE. A Si-rich crust is observed. **k-o**: Si-BE, BE-air and Si-air interfaces. Si-BE interface has barely detectable O content. Residues of the Si-rich crust are visible at the BE sidewall, similarly to the TE.

The fabricated Josephson junction barriers have a typical thickness of 2-3 nm (Fig. 3a), as is also commonly observed for angle-evaporated junctions [57]. No structural damage is seen on top of the BE due to Ar milling (Fig. 3a,b). However, traces of Ar can be detected at the barrier, which extends ~10 nm into the BE electrode (Fig. 3e). Similarly, Si atom traces can be found at the barrier, likely originating from the resputtered Si during the Ar milling (Fig. 3d). An atomic concentration of less than 4% for both elements is observed inside the junction (Supplementary Fig. 6). Impurities can lead to two-level system (TLS) defect losses in the junction [55,56,58]. In our experiment however, they cannot be distinguished from other residual junction losses. High resolution TEM images of overlap junctions fabricated with dry and wet etch processes show no discernible difference (Supplementary Fig. 4).

At the Si-TE interface, an amorphous Si layer of approximately 5 nm is visible (Fig. 1a: green interface, Fig. 3f). This is caused by Ar milling, which has been shown to damage the Si substrate [42]. This is also in agreement with traces of Ar found in this layer (Fig. 3j). Damaged Si layer has been reported to host TLS defects which lead to microwave losses [42]. At the Si-air interface however, no amorphous Si layer can be observed (Fig. 1a: yellow interface, Fig. 3g, l) and no Ar can be detected (Fig. 3j,o). We conclude that the amorphous Si is consumed by the TE etch, except below the TE. The potentially lossy amorphous Si layer constitutes only a small fraction of the total qubit footprint (Supplementary Fig. 5) and likely contributes to the residual junction losses. Near the TE sidewalls, we observed crust residues (Fig. 3g and Supplementary Fig. 4) that consist of oxygen (O) (Fig. 3h) and Si atoms (Fig. 3i). This crust arises from Si deposition on the sidewall and top of the resist during subtractive dry etching [59] (Fig. 1a: red interface). Similar to the amorphous Si layer, this crust can lead to residual microwave losses, however, only in the area surrounding the TE.

A less than 1 nm thin substrate-BE interface layer (Fig. 1a: orange interface, Fig. 3k) shows barely detectable traces of O impurities (Fig. 3m). This indicates that the substrate surface passivation with HF prior to the BE metal deposition successfully prevents oxide growth. The BE shows a distinct metal-air interface profile (Fig. 1a: brown interface), which we attribute to a combination of Ar milling, TE etching and residual BE crusts in case of BE dry etching. Residues of the BE crust and TE Al may be found at the sidewalls due to the anisotropic nature of the Ar milling and TE dry etch (Fig. 3l-n). The absence of Ar signal on the BE-air interface (Fig. 3o) indicates that the TE etch removes the amorphized layer from both the BE-air and Si-air interface.

## Discussion

As shown above, qubit devices are limited by the losses associated with the qubit capacitor pads. These losses can originate from crust residues, Al sidewall residues and native oxides near the BE structures. Since coherence times of wet and dry etched qubits do not significantly differ (Fig. 2c), we can exclude the BE crust remnants from the dominant loss sources. Other contributions cannot be confidently separated at this stage. However, advanced surface treatment steps can be devised to reduce its effects in future implementations.

Residual losses associated with the Josephson junction encompass loss from (i) quasiparticle tunnelling, (ii) the stray junction, (iii) crusts and native oxides in the vicinity of the junction, (iv) the amorphous Si interface underneath the TE, and (v) the junction barrier. Quasiparticle tunnelling losses can be alleviated by proper filtering of infrared [60] and high-energy radiation [61] or including quasiparticle trapping sites on chip [62]. Stray junction loss can be avoided by using a bandage process to ensure a proper galvanic contact between the TE and capacitor pad [15,42]. Similarly, as discussed for capacitor losses, crust and native oxides in the vicinity of the junction could be further reduced by developing dedicated surface cleaning processes. On the other hand, the amorphous Si and junction impurities are loss sources specific to Ar milling and can therefore not be easily eliminated. The effect of amorphous Si below the TE could be reduced by minimizing the TE surface area. Potential losses from junction contaminations can only be alleviated by further optimizing the Ar milling step. However, within our measurement uncertainty, their effect currently cannot be distinguished from other losses. To further investigate the junction losses, different qubit designs, such as 3D transmons [51], merged-element transmons [55] or finMETs [32] could be explored.

## Conclusions

We have fabricated and measured record-high coherence times for qubits based on overlap junctions, with mean $T_1$ values comparable to qubits fabricated with double-angle evaporated junctions. We demonstrated a qubit fabrication process that is fully compatible with CMOS fabrication requirements and tools. By studying different qubit designs, we infer that qubit limiting losses are located at the capacitor surface interfaces, and that our overlap junctions are not limiting the qubit relaxation times up to at least 103 μs. Further advancements in qubit lifetimes could be achieved by developing specialized surface treatments. The presented qubit fabrication process paves the way for reproducible and well controlled qubit integration in 300 mm wafer fabrication facilities, expediting the upscaling of quantum computers and other superconducting Josephson junction devices.

## Methods

### Overlap Josephson junction fabrication process

The substrate coupons for the qubit fabrication process are diced from two types of intrinsic Si wafers: 300 mm Si wafer with resistivity 3,000 Ω.cm and 100 mm Si wafer with resistivity >20,000 Ω.cm labelled 3k and 20k, respectively. After a close coupled wet etch of the native Si oxide with hydrofluoric acid, a 70 nm of Al is deposited by sputtering. This layer is used to form the qubit control and readout structures (BE layer). A 300 mm industry grade e-beam tool is used to define structures in the PMMA resist layer. The pattern is transferred to the BE layer either by dry or wet etch. The wet etch process is based on TMAH. The dry etch process is Cl-based and forms a shallow recess into the Si substrate. During dry etching in the presence of PMMA resist, a Si rich sidewall crusts forms on the BE and resist layer sidewalls. A dedicated solvent is used to simultaneously remove the resist and minimize sidewall crusts. After resist strip, the native Al oxide on top of the BE is removed by a coupon-wide Ar milling step in the load lock of the sputtering system. The Josephson junction (JJ) barrier layer is formed by controlled dynamic oxidation at constant pressure. Without vacuum break, 50 nm of Al is deposited to form the top electrode (TE). The subsequent TE is patterned using dry etching. The full process flow is visualized in Supplementary Fig. 1. A list of all fabricated samples in this study is found in Supplementary Table 1.

**Material characterization**

TEM, STEM and EDS measurements were performed at the Materials Characterization and Analysis centre at the IMEC microelectronics research institute. Before characterization, the samples were coated with spin-on carbon (SOC) layer. Lamellae with thickness of <50 nm were cut with focused ion beam (FIB) using Helios 450. TEM, STEM, and EDS measurements were performed on Titan Cubed Themis 300 STEM with a 200 kV source. For this study, TEM, EDS, atomic-resolution HAADF-STEM and ABF-STEM were used to investigate interfaces between substrate-top electrode, substrate-bottom electrode and the $AlO_x$ barrier between the top and bottom Al electrodes.

**Qubit measurement setup**

Overlap qubits are measured with a standard dilution-refrigerator setup (Supplementary Fig. 3). Input lines are thermalized with 20 dB attenuators at three different temperature stages. High frequency noise above the measurement frequency range is filtered before reaching the qubit with a low-pass filter (VLF-8400+) with cut-off frequency of 8.4 GHz. Each superconducting qubit chip is placed and wire bonded to the non-magnetic gold-plated copper PCB enclosed in an O-free copper sample holder. The sample holder is thermalized to the mixing chamber plate in a dilution refrigerator. The sample holder is surrounded by the copper radiation shield as well as two cryo-perm shields to minimize magnetic field at the sample. Output signal lines are thermalized with three isolators (LNF-ISC4_8A) with a total reverse isolation of ~ 60 dB and a 4-8 GHz band-pass filter (KBF-4/8-2S). The signal is amplified with a HEMT amplifier (LNF-LNC4_8C) at the 4K stage and ultra-low noise amplifier (LNA-30-04000800-07-10P) at room temperature.

At room temperature, pulsed signals are generated and acquired using the Keysight Quantum Engineering Toolbox: M3202A AWGs and M3102A Digitizer. The qubit excitation and readout pulse are combined at room-temperature and applied to the qubit's feedline. No dedicated charge line or flux line were used to excite or bias qubits.

Three qubit designs were used in this study, two Xmon (XM1, XM2) [50] styles and a two-pad transmon (TM) [14] qubit. Qubit frequencies were targeted in the range of 3-4 GHz and readout resonator frequencies are in the range of 5-6 GHz. Qubit anharmonicities are between 200 and 250 MHz.

**Numerical Simulations of the Participation Ratios**

The electric field losses are dependent on and are proportional to the participation ratios of the dielectric regions in the devices. The participation ratio $p_i$ in a dielectric $i$ is defined as the fraction of electrical energy contained in the dielectric with respect to the total electrical energy in the device, as follows [63]:

$$p_i = \frac{\epsilon_i \int_{V_i} |E_V|^2 dV}{\sum_i \epsilon_i \int_{V_i} |E_V|^2 dV}, \qquad (1)$$

where $\epsilon_i$ and $V_i$ correspond to the dielectric constant and volume of the $i^{th}$ dielectric respectively, and $|E_V|$ correspond to the absolute value of the electric field in the dielectric in an incremental volume $dV$. To calculate the participation ratio's $p_i$, we first estimate the electric fields by solving the Poisson's equation in the devices using a commercial TCAD electrostatic simulator [64], and then subsequently incorporate the fields into equation (1). We invoke the following approximations for calculating the electric fields in TCAD. First, we omit the Josephson junction region in the electrostatic model, as we are primarily interested in the losses in the capacitor region of the superconducting qubits. Second, with the geometry of the capacitor being largely symmetric along certain axes (see Supplementary Fig. 5a), we perform a two-dimensional electrostatic simulation, rather than a full 3-dimensional calculation. This further offers the advantage of reduced computational time and larger numerical accuracies for the simulations with an extremely refined mesh at interfaces in the device. Third, we only consider metal-air (MA) and substrate-air (SA) oxide interfaces in the simulations, while omitting the metal-substrate (MS) oxide interface. This is also reasonable considering that there is no substantial oxide formation in between the metal and substrate, as indicated in Fig. 3c and Fig. 3m. Furthermore, uncertain parameters are the exact thickness of the oxide at the different interfaces and their dielectric constant. From TEM images, we estimate an average thickness of the SA and MA interfaces of 4 and 5 nm, respectively. The geometry of the structures used in the simulations is very similar to that of the experimental devices, and we also employ commonly chosen values for the dielectric constants i.e., 11.9, 3.9 and 5.0 for the silicon substrate, and silicon dioxide and metal-air oxide interfaces respectively.

Based on the above technique, approximations, and parameters, we estimate the participation ratios in the capacitor regions for the three designs (resonator, Xmon and transmon) and shown in Supplementary Table 2. These obtained participation ratios have further been used to illustrate the scaling trend of $1/Q$ for the four devices in Fig 2d. The linear scaling with the participation ratio calculations in Fig 2d indeed confirms that the major limiting region for the qubit or resonator lifetimes ($T_1$) is the capacitor.

**Randomized benchmarking: measurement and analysis**

The average physical gate error $r_g$ and gate fidelity $F_{1q}$ have been measured for device C2 (Fig. 1d). A random sequence of Clifford gates of varying lengths $n$ is applied to a qubit initialized in the ground state. At the end of the sequence, an inverting gate is added to create an overall identity operation and the final qubit state is measured to determine the fidelity. The measurement is repeated 80 times for each sequence length. Cosine pulses with a duration of 50 ns are used in the experiment. The averaged sequence fidelity is fitted to $F = Ap^n + B$, where the parameters A and B depend on the state preparation and measurement errors [43]. From $p$, we determine $r_g$ as the error per Clifford $r_{Clifford}$ normalized by the average number of physical Clifford generator gate (1.875):

$$r_g = \frac{r_{Clifford}}{1.875} = \frac{(1-p)(d-1)}{1.875\,d}. \qquad (2)$$

Where $d = 2^m$ is the dimensionality of the Hilbert space, which is equal to 2 for a single qubit. The average physical gate fidelity is obtained from:

$$F_{1q} = 1 - r_g. \qquad (3)$$

The coherence limit on the measured fidelity ($F_{1q,inc}$) is estimated with qubit parameters: $T_1$ = 50 μs and $T_2^*$ = 60 μs [65].

**Data availability**

The data are available upon reasonable request.

**Code availability**

The codes are available upon reasonable request.

## Acknowledgments

The authors gratefully thank Kristiaan De Greve and Jonas Bylander for their comments after critical reading of the manuscript. The authors also gratefully thank Paola Favia, Hugo Bender and Chris Drijbooms for metrology support. This work was supported in part by the imec Industrial Affiliation Program on Quantum Computing.

## Author information

### Affiliations
**KU Leuven, Kasteelpark Arenberg 10, Leuven, B-3001, Belgium**
J. Verjauw, R. Acharya, J. Van Damme, D. Perez Lozano & M. Heyns

**Imec, Kapeldreef 75, Leuven, B-3001, Belgium**
J. Verjauw, R. Acharya, J. Van Damme, Ts. Ivanov, D. Perez Lozano, F. Mohiyaddin, D. Wan, J. Jussot, A. M. Vadiraj, M. Mongillo, M. Heyns, A. Potočnik, I. Radu, B. Govoreanu

### Author contributions
J.V. and A.P. planned the experiment. A.P. designed the samples. Ts. I., J.J., and D.L.P. preformed the sample fabrication, with contributions from D.W. Qubit data was collected by J.V., R.A. and A.M.V., and analysed by J.V. J.V.D. performed the resonator measurements. A.P. performed the randomized benchmarking and supervised the qubit characterization. Ts. I. performed the room temperature physical and electrical characterization. F.A.M. performed the participation ratio simulations. J.V. and A.P prepared the manuscript, with input from all authors. M.H, M.M., I.R., B.G., and A.P. supervised and coordinated the project.

### Corresponding authors

Correspondence to A. Potočnik [anton.potocnik@imec.be]

## Competing interests
The authors declare no competing interests.

# Supplementary information

**Supplementary Table 1**: Summary of measured devices. Devices include qubits and high-Q resonators. Devices located on the same sample are marked by a single letter. Roman numerals denote simultaneously fabricated samples (from the same die). Errors on reported average $T_1$ values represent the population's standard deviation (or standard deviation from the respective $T_1$ fit in case of single measurements). Junction areas have been measured for a specific device when values are marked by *. Other junction areas are estimated from test structures SEM images with the same design area.

| Device, die | Qubit geometry | BE etch | Wafer resistivity (Ω·cm) | Average $T_1$ (μs) | Device frequency (GHz) | Measurement span (days) | Measurement count | Junction area (μm²) |
|---|---|---|---|---|---|---|---|---|
| A1, I | TM | wet | 20k | 64.8±9.4 | 2.974 | 18.7 | 800 | 0.0136* |
| A2, I | TM | wet | 20k | 52.5±14.1 | 2.255 | 0.6 | 457 | 0.0142 |
| B1, II | TM | dry | 20k | 63.9±8.9 | 2.913 | 4.0 | 375 | 0.0496* |
| C1, II | TM | dry | 20k | 53.9±3.8 | 3.857 | 0.3 | 303 | 0.0751 |
| C2, II | TM | dry | 20k | 51.0±10.4 | 3.546 | 7.5 | 22 | 0.0643 |
| D1, III | TM | dry | 3k | 58.0±13.2 | 7.147 | 0.4 | 218 | 0.0780 |
| E1, III | TM | dry | 3k | 48.0±4.7 | 3.987 | 0.5 | 335 | 0.0868 |
| E2, III | TM | dry | 3k | 44.7±3.1 | 3.293 | 0.2 | 306 | 0.0528 |
| F1, II | XM2 | dry | 20k | 30.9±3.1 | 3.139 | 0.0 | 4 | 0.0326 |
| F2, II | XM2 | dry | 20k | 27.5±0.0 | 3.423 | 0.0 | 1 | 0.0407 |
| F3, II | XM2 | dry | 20k | 31.0±3.7 | 3.556 | 10.4 | 493 | 0.0495 |
| G1, I | XM1 | wet | 20k | 22.1±2.1 | 3.207 | 0.2 | 201 | 0.0091 |
| G2, I | XM1 | wet | 20k | 20.0±3.3 | 3.140 | 16.3 | 21 | 0.0284 |
| H1, II | XM1 | dry | 20k | 17.7±1.7 | 3.565 | 0.1 | 303 | 0.0407 |
| H2, II | XM1 | dry | 20k | 17.1±1.7 | 3.76 | 0.2 | 310 | 0.0495 |
| AR, I | RES | wet | 20k | 5.6±0.2 | 5.981 | 0.0 | 2 | - |
| BR, II | RES | dry | 20k | 5.5±0.4 | 5.889 | 0.0 | 3 | - |
| CR, II | RES | dry | 20k | 8.7±0.0 | 5.909 | 0.0 | 1 | - |
| DR, III | RES | dry | 3k | 7.4±0.0 | 5.76 | 0.0 | 1 | - |

**Supplementary Table 2**: Simulated participation ratios for the different dielectrics considered in the simulation. Note that the participation ratios in the metal oxide and SiO$_2$ dielectrics reduce with increasing gap $G$ between the conductors, with the transmon qubit having the smallest participation ratio in the oxides.

| Material | RES ($G$ = 4.5 µm) | XM1 ($G$ = 13 µm) | XM2 ($G$ = 24 µm) | TM ($G$ = 70 µm) |
|---|---|---|---|---|
| Silicon Substrate | ~0.91 | ~0.91 | ~0.92 | ~0.94 |
| Air | ~0.09 | ~0.09 | ~0.08 | ~0.06 |
| Metal-air | 7.5×10$^{-5}$ | 3.8×10$^{-5}$ | 2.2×10$^{-5}$ | 7.1×10$^{-6}$ |
| Substrate-air | 5.7×10$^{-4}$ | 2.8×10$^{-4}$ | 1.9×10$^{-4}$ | 5.8×10$^{-5}$ |

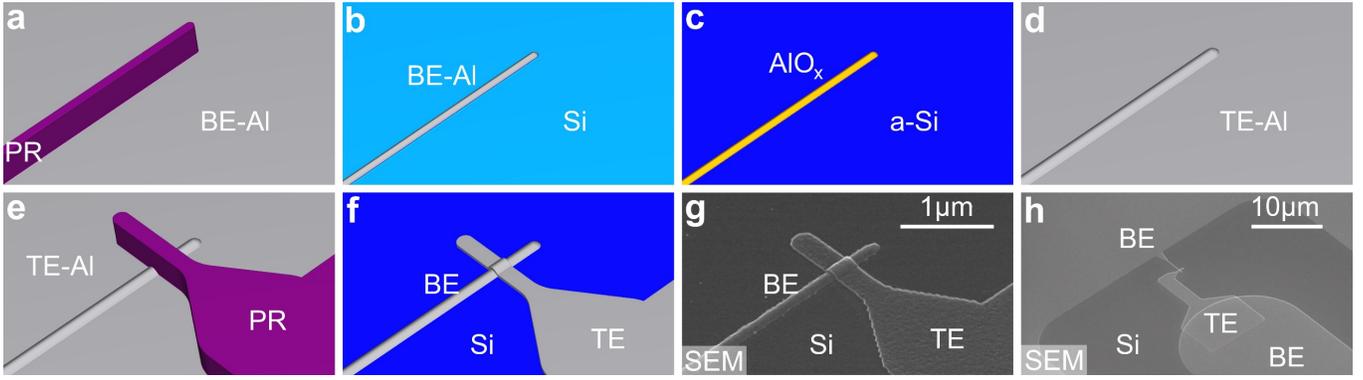

**Supplementary Figure 1**: Fabrication process for overlap Josephson junctions. **a** Photoresist (PR) is deposited and patterned on a 70 nm Al film which was sputtered on a high-resistivity Si wafer. **b** PR pattern is transferred to the Al layer by dry or wet etching to form the bottom electrode (BE). **c** Wafer is transferred to a deposition tool where the native oxides are first removed with Ar milling, followed by Josephson junction (JJ) barrier oxide growth with controlled dynamic oxidation. At the end of this process, the BE is covered by ~2 nm of $AlO_x$, together with an amorphized and oxidized Si surface (see Fig. 1b). **d** The second 50 nm thick Al layer is deposited in-situ on top of the BE. **e** PR is deposited and patterned on top of the second Al layer. **f** Using a dry etch, the PR pattern defines the top electrode (TE). **g** SEM image focusing on the overlap Josephson junction and **h** a zoomed-out SEM micrograph of a fully fabricated device, showing the large BE-TE spurious overlap junction.

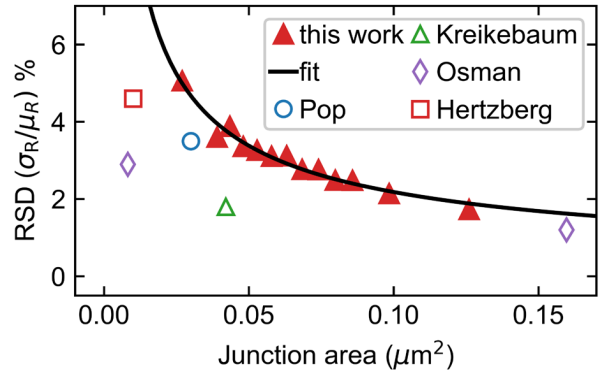

**Supplementary Figure 2:** Chip-level relative standard deviation (RSD) of room-temperature JJ resistances as a function of overlap junction area. Data is shown for test structures located on die III. For each RSD datapoint, at least 30 nominally identical JJ were measured. State-of-the-art chip-level RSD values of double-angle evaporated Josephson junctions adapted from literature are shown with empty symbols for reference (Pop, *et al.* [66], Osman, *et al.* [24], Hertzberg, *et al.* [23], Kreikebaum, *et al.* [25]). We note the overlap JJ resistance variation in our work is comparable to those of double angle JJ. RSD for previous reported overlap JJ is estimated to be notably higher (RSD = ~21% for JJ area: 0.0324 µm² [34]). Black line shows a fit to the measured RSD data using the following equation [24]: $RSD = \sqrt{a/A^\gamma + b}$, where $A$ is the junction area, $\gamma$ is a phenomenological exponent close to 1, $b$ is related to the barrier thickness variation and $a$ is a combination of junction thickness and area variation. The fitting procedure yields $b$ very close to 0, indicating that JJ resistance RSD is dominated by the area variation. We have expanded equation (3) in [24] with the phenomenological exponent $\gamma$ to account for junction area deviations from the SEM determined junction area. The resulting $\gamma = 1.3 \pm 0.1$ points to weak junction geometry dependence on the junction area.

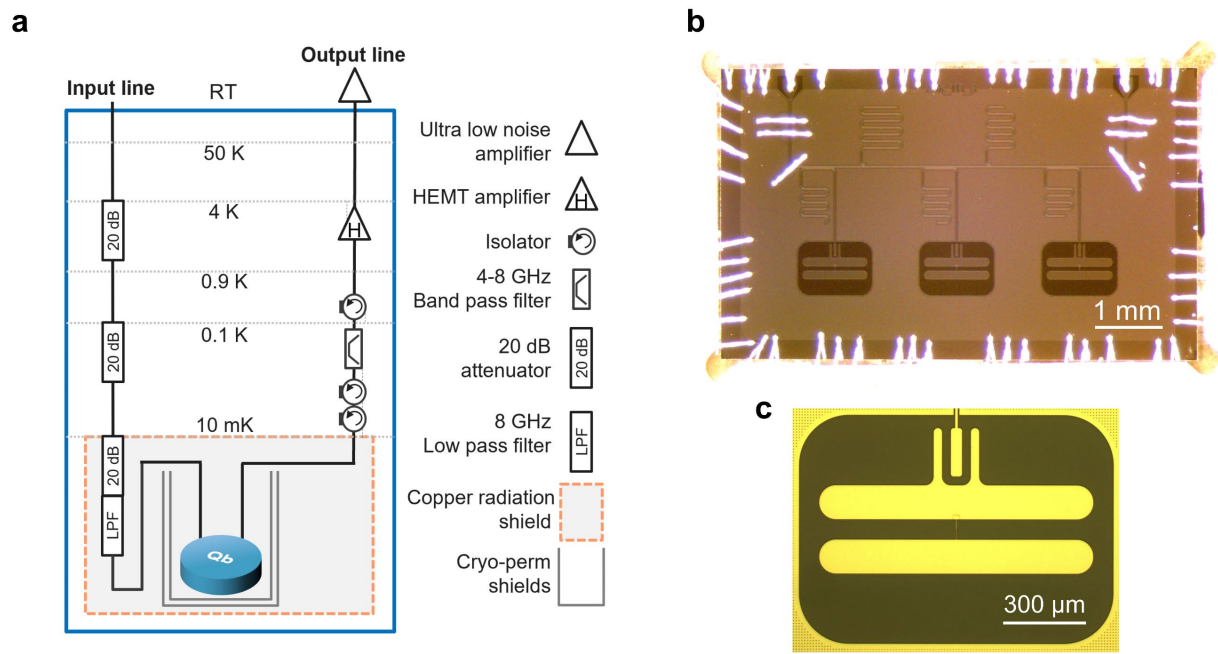

**Supplementary Figure 3:** Experimental setup illustration and device image. **a** Dilution refrigerator schematic for superconducting qubit characterization. **b** Image of a double-pad transmon qubit chip, containing three qubits. **c** Zoom-in of a single double-pad transmon qubit from the qubit chip presented in **b**.

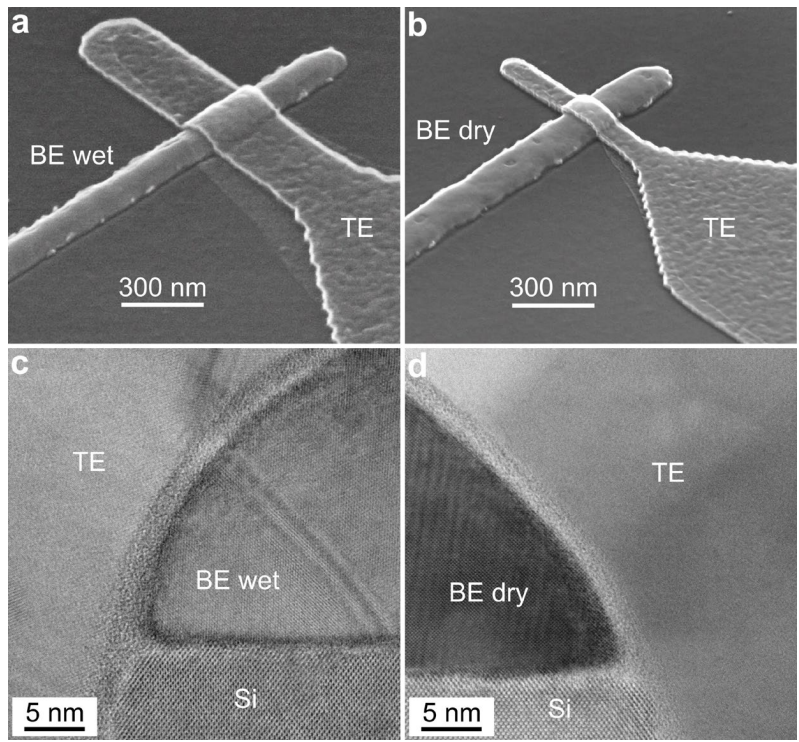

**Supplementary Figure 4**: Post fabrication etch comparison. **a** Wet and **b** dry etched bottom electrode SEM images from test devices in sample A1 and B1, respectively. In both cases the bottom electrode top surface and sidewall roughness is affected by the subsequent Ar milling step. From these images no qualitative morphological differences can be observed. A thin sidewall residue can be observed surrounding the top electrode. These come from crust formation during TE

etching. Based on EDS results presented in Fig. 3 in the main text, the residues are mostly composed out of oxidized Si. Similarly, high resolution ABF-STEM overlap junction images of **c** wet and **d** dry etched samples show that both processes yield virtually indistinguishable junction barriers. Both wet and dry processes result in comparable device performance as shown in the main text for qubits (Fig. 2c,d) and resonators (Fig. 2d).

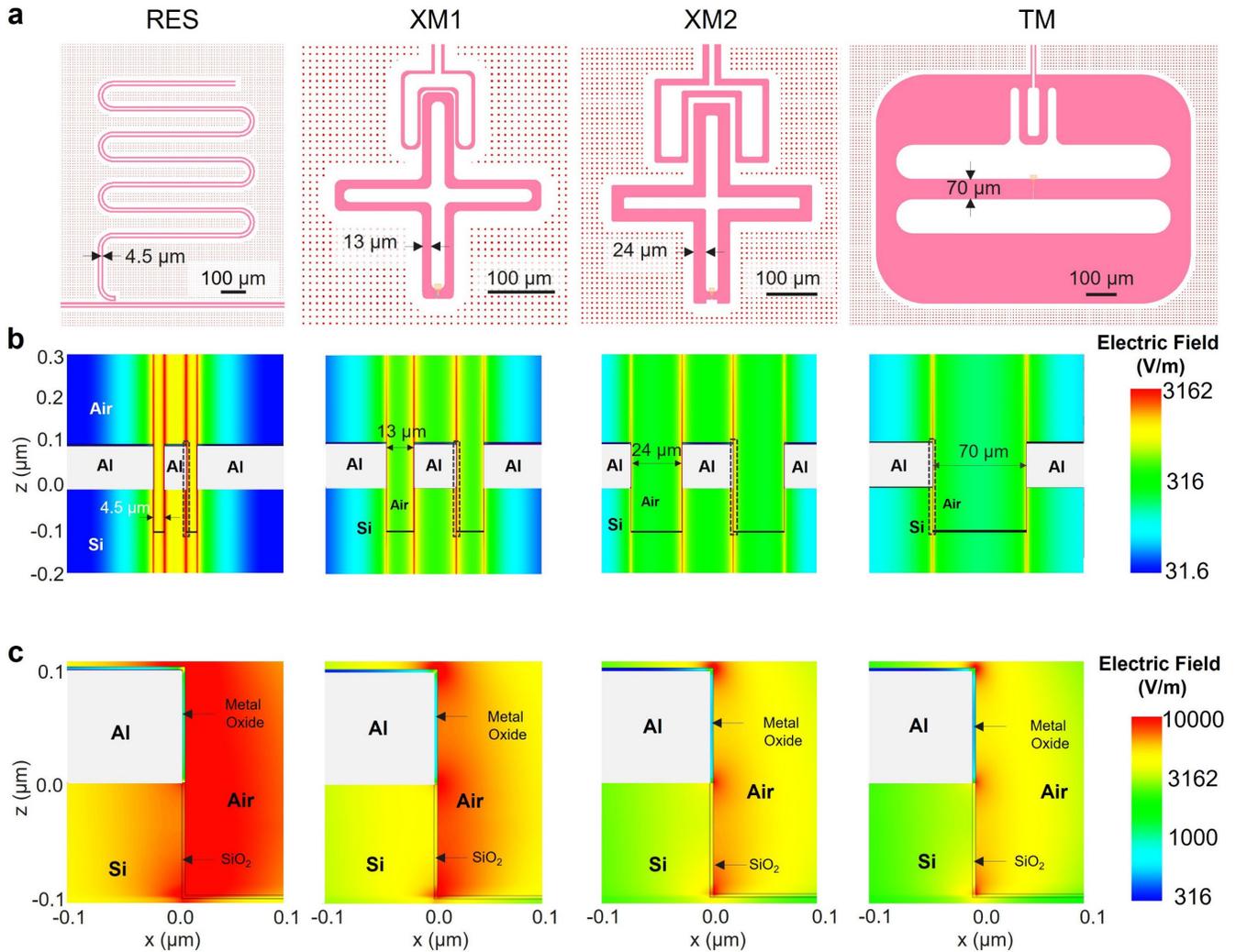

**Supplementary Figure 5:** Device layouts and simulation results **a** Two-dimensional representation of devices measured in the experiment including a coplanar waveguide resonator (RES), Xmon qubits XM1 and XM2 with different dimensions, and a transmon qubit TM. Devices are labelled and arranged from left to right in the order of increasing gap $G$. **b, c** Absolute electric field along a cross section of the four devices estimated using TCAD techniques. The electric fields over a wide region are shown in subfigure **b**, while subfigure **c** illustrates the fields in the vicinity of the substrate-air and metal-air interfaces, where different dielectric regions are labelled as well. Note that the electric fields in the dielectric regions of the devices decrease with increasing gap $G$. The colormap has been truncated in subfigure **b** and subfigure **c** for visual clarity when comparing the devices.

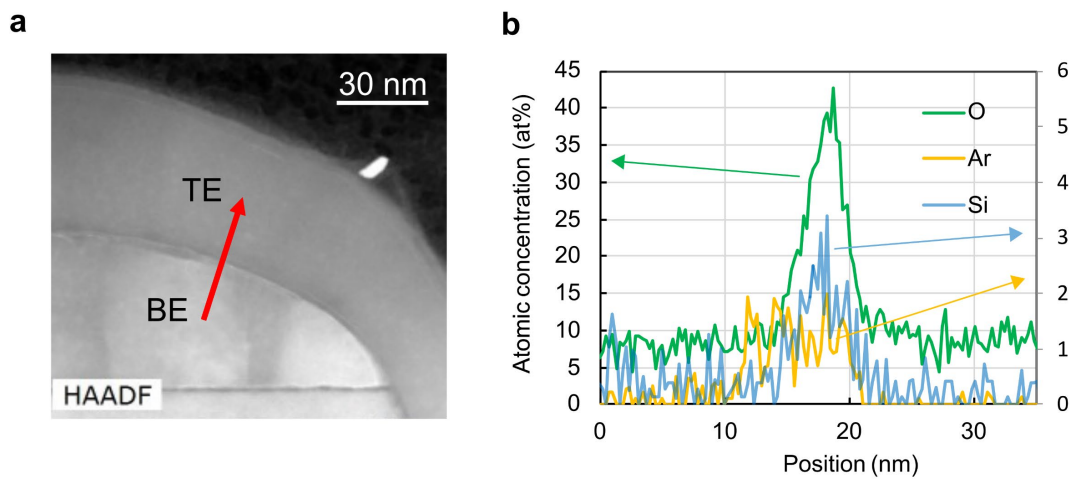

**Supplementary Figure 6**: Junction barrier trace element composition **a** Josephson junction TEM image (presented in Fig. 3. b in the main text). Linecut is indicated in red, along which the atomic concentration profiles are determined. **b** Atomic concentration profiles of the Josephson junction barrier. The position coordinate on the x-axis runs from BE to TE. The O (green) atomic concentration is shown on the left y-axis and its baseline content is attributed to EDS sample preparation. The barrier contains Ar (yellow) and Si (blue) trace elements along the linecut, respectively. Their atomic concentrations are shown on the right y-axis.